\documentstyle[12pt]{article}
\begin{document}
\def \P{ {P(\delta x,t)}}

\begin{center}
\centering{\bf Herd behavior and aggregate fluctuations in financial markets }
\footnote{This work is a condensed version of Chapter 5 the first author's doctoral dissertation
at Universit\'e de Paris XI. 
R. Cont gratefully acknowledges an AMX fellowship from Ecole Polytechnique (France)
and thanks Science \& Finance for their hospitality.
We thank Yann Braouezec for helpful remarks on a preliminary draft of this article and
for numerous bibliographical indications. E-mail:  cont@ens.fr}
\end{center}

\begin{center}
\centering{ Rama CONT $^{1,2,3}$ and Jean-Philippe BOUCHAUD$^{1,2}$ \\
\vskip 1cm

{\it 1) Service de Physique de l'Etat Condens\'e\\
 Centre d'Etudes de Saclay\\
91191 Gif sur Yvette , France}\\

{\it 2) Science \& Finance Research Group\\
 109-111 rue V. Hugo, 92 532 Levallois, France. }

{\it 3) Department of Economics\\
American University \\
31 avenue Bosquet 75007 Paris, France}

} 
\end{center}

\begin{abstract} 
We present a simple model of  a stock market where a random communication
structure between agents  generically gives rise to a heavy tails in the 
distribution
of  stock price variations in the form of an exponentially truncated power-law,
similar to distributions observed in recent empirical studies of high 
frequency market data.
Our model provides a link between two  well-known  market phenomena:
the heavy tails observed in the distribution of stock market returns
on one hand and 'herding' behavior in financial markets on the other hand.
In particular, our study suggests a relation between the excess kurtosis observed in
asset returns, the market order flow and the tendency of market participants
to imitate each other.
\end{abstract} 
 


\newpage

 Empirical studies of the fluctuations in the price of various financial assets
have shown that  distributions of stock returns and stock price changes
have fat tails that deviate from the Gaussian distribution
     \cite{mandelbrot63,mandelbrot97,campbell,olsen2,olsen,houches,pagan} 
especially for intraday time scales      \cite{houches}.
These fat tails, characterized by a significant  excess kurtosis,
persist even after accounting for heteroskedasticity in the data      \cite{bollerslev}.
The heavy tails observed in these distributions correspond to large fluctuations
in prices, "bursts" of volatility which are difficult to explain only in
terms of variations in fundamental economic variables      \cite{shiller1}.

The fact that significative fluctuations in prices are not necesarily related
to the arrival of information   \cite{cutler89} or
to variations in fundamental economic variables      \cite{shiller1}
leads to think
the high variability present in stock market returns may correspond 
to collective  phenomena such as crowd effects or "herd" behavior.

Although herding
in financial markets is by now relatively well documented empirically, there have been few
theoretical studies on the implications of herding and imitation for
 the statistical 
properties of market demand and price fluctuations. In particular some questions
one would like to answer is: how does the presence of herding modify the distribution of
returns? 
What are the implications of herding for relations between market variables such
as order flow and price variability?
These are some of the questions which have motivated our study.

The aim of the present study is to examine,
 in the framework of a simple model, how  the existence of herd behavior among market
participants may generically lead to large fluctuations in the aggregate excess demand, 
described by a heavy-tailed non-Gaussian distribution. Furthermore we explore how 
empirically measurable quantities such as the excess kurtosis of returns and the average
order flow may be related to each other in the context of our model. Our approach
 provides a quantitative link between the two   issues discussed above:
the {\it heavy tails} observed in the distribution of stock market returns
on one hand and the {\it herd } behavior observed in financial markets on the other hand.

\noindent
The article is divided into four sections. Section 1  reviews well known empirical facts about
the heavy-tailed nature of the distribution of stock returns and various models proposed
to account for it. Section 2 presents previous empirical and theoretical
work on herding and imitation in financial markets in relation to the present study.
 Section 3  discusses the statistical properties of excess demand resulting from the aggregation of 
of a large number of random individual demands in a market. Section 4 defines our model and presents
analytical results. Section 5 interprets the results in economic terms, compares them to empirical
data and discusses possible extensions. Details of calculations are given in the appendices.

\section{The heavy-tailed nature of asset return distributions}

It is by now a well known fact that the distribution of returns of almost all financial assets -stocks,
indexes and futures-  exhibit a slow asymptotic decay that deviates from a normal distribution. This
is quantitatively reflected in the excess kurtosis, defined as:
\begin{eqnarray}
\kappa &=& \frac{\mu_4}{\sigma^4} - 3 
\end{eqnarray}
where $\mu_4$ is the fourth central moment and $\sigma$ the standard deviation of the returns.
$\kappa$ should be zero for a normal distribution
but ranges between 2 and 50 for daily returns      \cite{campbell,pagan} and is even higher
 for intraday
data. Careful study of the tails of the distribution shows an exponential decay for most assets
     \cite{houches,stanley}. 

Many statistical models  have been put forth 
to account for the heavy tails observed in
the distribution of asset returns. 
Well known examples are 
Mandelbrot's stable paretian hypothesis      \cite{mandelbrot63}
the mixture of distributions hypothesis      \cite{clark},
and  models based on conditional heteroskedasticity      \cite{arch}.

It is well known that in the presence of heteroskedasticity,
the unconditional distribution of returns will have heavy tails.
In most models based on heteroskedasticity, the 
process of return is assumed to be conditionally Gaussian : the 
shocks are "locally" Gaussian and the non-Gaussian character of
the unconditional distribution is an effect of aggregation. It 
is obtained by superposing
a large number of local Gaussian shocks.
In this description, sudden movements in prices  are interpreted as 
corresponding to a high value of conditional variance.

On one hand, it has been shown that although conditional heteroskedasticity
does lead to fat-tails in unconditional distributions, ARCH-type models cannot
fully account for the kurtosis of returns      \cite{hsieh3,bollerslev}.
On the other hand, from a theoretical point of view 
there is no {\it a priori } reason to postulate that returns
are conditionally normal: although conditional normality is convenient for parameter
estimation of the resulting model, non-normal conditional distributions possess the same
qualitative features as for volatility clustering while accounting better for heavy tails.
Gallant and Tauchen      \cite{gallant} report  significant evidence of both
conditional heteroskedasticity and conditional non-normality in the daily NYSE
value-weighted index. Similarly, Engle and Gonzalez-Rivera      \cite{engle91}
show that when a GARCH(1,1) model is used for the conditional variance of stock 
returns the conditional distribution has considerable kurtosis, especially for small firm stocks. Indeed several authors have proposed GARCH-type models
with non-normal conditional distributions      \cite{bollerslev}.

Stable distributions      \cite{mandelbrot63} offer an elegant alternative to heteroskedasticity for
generating fat tails, with the advantage that they have a natural interpretation in terms
of aggregation of a large number of individual contributions of agents to market fluctuations:
indeed,  stable distribution may be obtained as limit  distributions of sums of independent
or  weakly dependent  random variables, a property which is not shared by alternative models.
Unfortunately, the infinite variance property of these distributions is not observed
in empirical data: sample variances do not increase indefinitely with sample size but appear
to stabilize at a certain value for large enough data sets.
We will discuss stable 
distributions in more detail in Section 3.

A third  approach, first advocated by Clark     \cite{clark}, 
is to model stock returns by a  subordinated process, typically 
subordinated Brownian motion. This amounts to stipulating that
through a ``stochastic time change" one can transform the complicated dynamics
of the price process into Brownian motion or some other simple process.
It can be shown that, depending on the choice
of the subordinator, one can obtain a wide range of  distributions for the increments
all of which  possess heavy tails i.e. positive excess kurtosis. 
As a matter of fact, even stable distributions
may  be obtained as a subordinated Brownian motion.
In the original approach of Clark     \cite{clark}, the subordinator was taken
to be trading volume. Other choices which have been proposed are the number
of trades      \cite{geman} or other local measures of market activity.
However, none of these choices  for the subordinator
lead to a normal distribution for the increments of the time-changed process,
indicating that  large fluctuations in price may not be completely explained by
large fluctuations in trading volume or number of trades.

In short, although heteroskedasticity and time deformation partly explain the 
kurtosis of asset returns, they do not explain it quantitatively: even after
accounting for these effects, one is left with an important residual kurtosis
in the resulting transformed time series. 
Moreover, these approaches
 are not based on any particular model of the market
phenomenon generating the data that they attempt to describe.

Recent works by Bak, Paczuski and Shubik     \cite{shubik} and 
Caldarelli, Marsili and Zhang      \cite{caldarelli}
have tried to explain the heavy tailed nature of return distributions
 as an emergent property in a
market where fundamentalist traders interact with noise traders.
Bak, Paczuski and Shubik consider several types of trading rules
and study the resulting statistical properties for
the time series of asset prices in each case.
Computer simulations of their model do seem to yield  fat-tailed distributions
for asset returns which at least qualitatively resemble empirical distributions of 
stock returns, showing that the appearance of fat-tailed distributions
can be regarded as an emergent property in large markets. 
However, the model has two drawbacks:
 first, it is a fairly complicated model with
many ingredients and parameters and it is difficult to see
how each ingredient of the model  affects the results obtained, which
in turn diminishes its explanatory power. Second, the complexity of the
model does not allow explicit calculations to be performed, 
preventing the model parameters to be compared with empirical values.

We present here an alternative approach which,  
by modeling the communication structure between market agents as a {\it random graph}, 
proposes a simple mechanism accounting for some non-trivial statistical properties
of  stock price fluctuations. Although much more rudimentary and 
containing fewer ingredients than the model proposed by Bak, Paczuski and
Shubik, our model allows for analytic calculations to be performed, thus
enabling us to interpret in economic terms the role of each of the parameters
introduced. The basic intuition behind our approach is simple: 
interaction of market participants through imitation can lead to 
large fluctuations in aggregate demand, leading to heavy tails in the distribution of returns.

\section{Herd behavior in financial markets}

A number of recent studies have  considered mimetic behavior as a 
possible explanation for the excessive volatility observed in financial markets
\cite{bannerjee93,shiller1,topol}.
The existence of herd behavior in speculative markets has been documented by a certain number
of studies: Scharfstein and Stein      \cite{scharfstein} discuss evidence of herding in the behavior
of fund managers, Grinblatt {\it et al.}      \cite{grinblatt} report 
herding in mutual fund behavior while Trueman      \cite{trueman} and Welch      \cite{welch} 
show evidence for herding in the forecasts made by financial
analysts. 

On the theoretical side several studies have shown that, in a market with noise traders,
herd behavior is not necessarily ``irrational" in the sense that it may be compatible
with optimizing behavior of the agents      \cite{shleifer}. Other motivations may be invoked
for explaining imitation in markets, such as ''group pressure"      \cite{bikhchandani,scharfstein}.

Various models of herd behavior have been considered in the literature, the most well 
known approach being that of Bannerjee      \cite{bannerjee,bannerjee93}
and  Bikhchandani {\it et al.}
     \cite{bikhchandani}. In these models, individuals attempt to infer a parameter from
noisy observations and decisions of other agents, typically through a Bayesian procedure,
 giving rise to "information cascades"      \cite{bikhchandani}.
 An important feature of these models is the sequential
character of the dynamics: individuals make their decisions 
one at a time, taking into account the decisions of the individuals preceding them.
The model therefore assumes a natural way of ordering the agents.
This assumption seems unrealistic in the case of financial markets: orders from various
market participants enter the market simultaneously and it is the interplay between different
orders that determines aggregate market variables\footnote{Bikhchandani {\it et al} 
     \cite{bikhchandani}
do not consider their model as applicable to financial markets but for another reason:
they remark that
as the herd grows, the cost of joining it will also grow, discouraging new agents to join.
This aspect, which is not taken into account by their model, is again unavoidable to the
sequential character of herd formation.}.

Non-sequential herding has been studied in a Bayesian setting by Orl\'ean      \cite{orlean}
in a framework inspired by  the Ising model. Orl\'ean considers imitation between agents
in which any two agents  have the same tendency to imitate each other.
 In terms of aggregate variables, this model leads either to a Gaussian distribution
when the imitation is weak, or to a bimodal distribution with non-zero modes, which Orl\'ean
interprets as corresponding to
collective market phenomena such as crashes or panics. 
In neither case does one obtain a heavy-tailed unimodal distribution centered at zero
such as those observed for stock returns.

The approach proposed in this paper is different from both approaches described above.
Our model is different from that in      \cite{bannerjee,bikhchandani}.
in that herding is not sequential.
The unrealistic nature of the results in      \cite{orlean} result from
the fact that all agents are assumed to imitate each other to the same degree.
We avoid this problem by considering the random formation of groups of agent
who imitate each other but such that different groups of agents make independent
decisions, which allows for a heterogeneous market structure. 
More specifically, our approach considers 
the interactions between agents as resulting from a 
{\it random} communication structure, as explained below.

\section{Aggregation of random individual demands}

 Consider a stock market with $N$ agents, labeled 
by an integer $1 \leq i \leq N$, trading in a single asset, whose price at time
$t$ will be denoted $x(t)$. During each time period, an agent may choose either to buy the stock, sell it or not to trade.
The demand for stock of agent $i$ is  represented by a random variable  $\phi_{i} $, which can take the values 0,  -1 or +1: a positive value  of $\phi_{i}$  represents a "bull"- an agent willing to buy stock-, a negative value a "bear", eager to sell stock while $\phi_i = 0$ means that agent $i$ does
not trade during a given period. The random character of individual demands may 
be due either to heterogeneous preferences or to random resources of the agents, or both.
 Alternatively, it may result
from the application by the agents of simple decision rules, each group of agents
using a certain rule.
However, in order to focus on the effect of herding,
 we do not explicitly model  the decision
process leading to the individual demands  and model the result of the decision 
process as a random variable $\phi_i$.  
In contrast with many binary choice models 
in the microeconomics literature, we allow for an agent to be inactive i.e. not to trade during 
a given time period $t$. This, as we shall see below, is  
important for deriving our
results.
 
Let us consider for simplification that, during each time period, an agent may either trade
 one unit of the asset or remain inactive. The demand of the agent $i$ is then  represented by
$\phi_{i} \in \{-1,0,+1\}$, $\phi_{i}=-1 $
representing a sell order. The aggregate excess demand for the asset at time $t$
is therefore
simply 

\begin{eqnarray}
 D(t) &=&  \sum _{i=1}^{N} \phi_{i}(t) \label{demand}
\end{eqnarray}

 given the algebraic nature of the  $\phi_{i}$. The marginal distribution of agent $i$s individual demand will is assumed to be symmetric :

$$
 P(\phi_i = +1) = P(\phi_i = -1) = a \qquad  P(\phi_i = 0) = 1-2a 
$$

such that the average aggregate excess demand is zero i.e. the  market is considered to fluctuate  
around equilibrium. 
A value of $a< 1/2$ allows for a finite fraction of agents not to trade
during a given period.

We are concerned here with obtaining a result which could then be
compared
with actual market data and 
the short term excess demand is not an easily observable quantity.
Also, most of the studies on the statistical properties of financial
time series have been done on returns, log returns or price changes.
We therefore need to relate the aggregate excess demand in a given
period to the return or price change during that period.
The aggregate excess demand has an impact on the price of the stock, causing it to
rise if the excess demand is positive, to fall if it is negative.
A common specification, which is compatible with standard {\it tatonnement}
ideas, is to assume a proportionality between price change (or return)
and excess  demand:

\begin{eqnarray}
{\Delta x} = x(t+1) - x(t)& =& \frac{1}{\lambda} \sum _{i=1}^{N} \phi_{i}(t)    \label{price}
\end{eqnarray}

where $\lambda$ is a factor measuring the liquidity or, more precisely, the {\it market depth}      \cite{kempf}: it
is  the excess demand needed to move the price by one unit: it measure the sensitivity of price to fluctuations
in demand. 
Eq. (\ref{price}), emphasizes the price impact of the order flow as opposed to other
possible causes for price fluctuations.
Eq.(\ref{price}) may be considered either in absolue terms with $x(t)$
being the price, or as representing {\it relative} variations
of the price, $x(t)$ then being considered as the log of the price
and its increment as the instantaneous return. The latter has the advantage
of guaranteeing the positivity of the price but for short-run dynamics
 the two specifications do not differ substantially since the two quantities
have the same empirical properties.
A similar model for the price impact of {\it trades} has been considered by
Hausman, Lo and MacKinlay      \cite{hausman}.
Although in the long run
economic factors other than short term excess demand 
may influence the evolution of the asset price, resulting
in mean-reversion or more complex types of behavior, we
focus here on the short-run behavior of
prices, for example on intra-day time scales in the 
case of stock markets, so this approximation seems reasonable.
The linear nature of this relation may also be questioned: 
indeed, some empirical studied seem to indicate that the price impact of
trades may be non-linear      \cite{campbell,kempf}. First, note that these studies 
deal with the price impact of trades and not of order flow (excess demand), which
is much harder to measure. Results reported by Farmer and co-workers      \cite{farmer} 
based on the study of the price impact of blocks of orders of different sizes sent to
the market seem to indicate a linear relationship for small price changes with
nonlinearity arising when the size of blocks is increased.
Moreover, if the one-period return $\Delta x$
is a non-linear but smooth function $h(D)$
\footnote{It is interesting to note that if ${\Delta x}=h(D)$, where
$h$ is an increasing function of $D$ and if the individual demands $(\phi_i(t))$
are sequences of 
independent random variables (a somewhat extreme assumption), then it is easy to show that the
overall wealth of all traders increases on average with time.} of the excess demand, then
a linearization of the inverse demand function $h$ (a first order
Taylor expansion in $D$) shows that Eq.(\ref{price}) still holds
for small fluctuations of the aggregate excess demand 
with $h'(0)= 1/\lambda$.

In
order to evaluate the distribution of stock returns from  Eq.(\ref{price}), we  need to know the {\it joint}
distribution of the individual demands $(\phi_i(t))_{1 \leq i\leq N}$. 
Let us begin by considering the simplest case where individual demands  $\phi_i$ of
 different agents are independent identically distributed random
variables. We shall refer to this hypothesis as the "independent agents" hypothesis.
In this case the joint distribution of the individual demands
is simply the product of individual distributions and 
the  price variation $\Delta x$ is a sum of $N$ {\it iid}
random variables with finite variance. When the number of terms in Eq.(\ref{price}) is large
the central limit theorem applied to
the sum in Eq.(\ref{price}) tells us that the distribution of $\Delta x$ is well approximated
by a Gaussian distribution. 
Of course, this result still holds as long as the distribution of individual demands 
has finite variance.

This can be seen as a rationale for the frequent use of the normal
distribution as a model for the distribution of stock returns: indeed,
if the variation of market price is seen as the sum of a large number of
independent or weakly dependent random effects, 
it is plausible that a Gaussian description should be 
a good one.

Unfortunately, empirical evidence tells us otherwise: the distributions
both of asset returns      \cite{pagan,campbell} and of asset price changes      \cite{mandelbrot63,mandelbrot97,houches,scaling} have been repeatedly shown to deviate significantly 
from the Gaussian distribution, exhibiting fat tails and excess kurtosis.

But the independent agent model is  also capable of generating aggregate distributions
with heavy tails: indeed, if one relaxes the assumption that  the individual demands 
$\phi_i$ have a finite variance then under 
the hypothesis of independence (or weak dependence) of individual demands, the aggregate
demand -and therefore the price change if we assume  Eq.(\ref{price})- will have 
a stable (Pareto-L\'evy) distribution. This is a possible interpretation for the stable-Paretian
model
 proposed by 
Mandelbrot      \cite{mandelbrot63} for the heavy tails observed in
the distribution of the increments of various market prices. 
The infinite variance of the $\phi_i$ then reflects  the heterogeneity of the market,
for example  in terms  of  broad distribution of 
wealth of the participants as proposed by
Levy \& Solomon \cite{solomon}.

Mandelbrot's stable-Paretian hypothesis has been criticized for 
several reasons, one of them being that
 it predicts an infinite variance for stock returns
which implies in practice that the sample variance will indefinitely increase
with sample size, a property which is not observed in empirical data.

More precisely, a careful study of the tails of the distribution of increments 
for various financial assets shows      \cite{book,houches} that they 
have heavy tails with a finite variance. Many distributions verify these conditions
     \cite{campbell}; a particular example  proposed by the authors and others     \cite{houches} 
is an exponentially truncated stable distribution 
the tails of the density then have the asymptotic form of  an exponentially truncated
power law:

\begin{eqnarray}
p(\Delta x) \mathop{\sim}_{|\Delta x|\to\infty} \frac{C}{|\Delta x|^{1+\mu}} exp \ (-\frac{\Delta x}{\Delta x_0})
\end{eqnarray}

The exponent $\mu$ is found to be close to 1.5 ($\mu \simeq 1.4 - 1.6$) for a wide variety of stocks and market indexes      \cite{book}.
This asymptotic form allows for heavy tails (excess kurtosis)
without implying infinite variance. \\

\noindent
However, it is known the central limit theorem also holds for  certain sequences of
 dependent variables:  under various types of  {\it mixing}
conditions      \cite{billingsley}, which are mathematical formulations of the notion of
``weak" dependence,  aggregate variables will still be normally distributed.
Therefore the non-Gaussian and more generally non-stable 
character of empirical distributions, be it excess demand or the stock returns,
 not only
demonstrates the failure of the ``independent agent" approach, but also shows that
such an approach is not anywhere close to being a good approximation: the dependence between
individual demands is an essential character of the market structure and may not be left out
in the aggregation procedure,
 they cannot be assumed to be
"weak" (in  the sense of a mixing condition      \cite{billingsley}) and {\it do}
change the  distribution of the resulting aggregate variable, 

Indeed, the assumption that the outcomes of decisions of individual agents
may be represented as independent random variables is highly unrealistic:
such an assumption ignores an essential ingredient of market organization,
namely the {\it interaction} and {\it  communication} among agents.

In real markets, agents may form groups
 of various sizes which then
may share information and act in coordination. In the context of a financial market,
groups of traders may align their decisions
and act in unison 
to buy or sell;
a different interpretation of a "group" may be 
an investment fund corresponding to the wealth of several investors 
but managed by a single fund manager. 

In order to capture such effects we need to introduce an additional ingredient, namely the communication structure between agents. 
One solution would be to specify a fixed trading group structure and then 
proceed to study the resulting aggregate fluctuations. Such 
an approach has two major drawbacks. 
First, a realistic market structure may require specifying
a complicated structure of clusters and rendering the resulting model
analytically intractable. More importantly, the resulting pattern of
aggregate fluctuations will crucially depend on the specification of
the market structure. 

An alternative approach, suggested by Kirman      \cite{kirman1},
 is to consider the market communication structure
itself as stochastic.
One way of generating  a random market structure is to
assume that market participants meet randomly and trades
take place when an agent willing to buy meets and agent willing to sell.
This procedure, called "random matching" by some authors     \cite{ioannides1},
has been previously considered in the context of formation of trading groups
by Ioannides      \cite{ioannides1} and in the context of a stock market model by Bak, Paczuski and Shubik
     \cite{shubik}. 
Another way  is to consider that market participants form groups or``clusters" 
through a random matching  process but that no trading takes places inside a given group: 
 instead, members of a given group adopt a common market strategy (for example, they decide to
buy or sell or not to trade) and different groups may trade with each other through
a centralized market
process. In the context of a financial market, clusters
may represent for example a  group of investors participating in a mutual fund.
This is the line we will follow in this paper.

\section{Presentation of the model}

More precisely, 
let us suppose that  agents group together in coalitions or {\it clusters}
 and that, once a coalition has formed, all it's members
coordinate their individual demands so that all individuals in a given
cluster have the same belief regarding future movements
of the asset price. In the framework described in the preceding section,
we will consider that all agents belonging to a given cluster will
 have the same demand $\phi_i$ for the stock. 
In the context of a stock market,
these clusters may correspond for example
to mutual funds e.g.  portfolios managed by the same fund manager or
to herding among security analysts as in      \cite{trueman,welch}.
The right hand side of the  equation (3) may therefore be rewritten as a sum over clusters :

\begin{eqnarray}
{\Delta x} = \frac{1}{\lambda} \sum _{\alpha=1}^{k} W_{\alpha} \phi_{\alpha}(t) &=&
\frac{1}{\lambda} \sum _{\alpha=1}^{n_c} X_{\alpha}\label{somme}
\end{eqnarray}

where $W_{\alpha}$ is the size of cluster $\alpha$, $\phi_{\alpha}(t)$ the
(common) individual demand of agents belonging to the cluster $\alpha$,
$n_c$ the number of clusters (coalitions) and $X_\alpha = \phi_\alpha W_\alpha$.
 
One may consider that coalitions 
are formed through binary links between agents, a link between two agents meaning that they undertake the same action on the market i.e. they both buy or sell stock.
 For any pair of agents $i$ and $j$, let $p_{ij}$
be the probability that $i$ and $j$ are linked together. Again, in order to simplify, we assume that $p_{ij} = p$ is independent  of $i$ and $j$: all links are equally probable. $(N-1) p$ then denotes the average
number of agents a given agent is linked to. Since we are interested in studying
the $N\to \infty$ limit, $p$ should therefore be chosen in such a way
that $(N-1) p$ has a finite limit. A natural choice is $p_{ij} = c/N$, any other
choice verifying the above condition being asymptotically equivalent to this one. 
The distribution of
coalition sizes in the market is thus completely specified  by 
a single parameter, $c$, which represents the  willingness of agents to align their actions: 
it can be interpreted as a coordination number,
measuring the degree of clustering among agents.

Such a structure is known as a {\it random graph} in the mathematical
literature      \cite{erdos,bollobas}:  in terms of random graph theory,
 we consider agents as vertices of a random graph of size $N$,
and the coalitions  as connected components of the graph.
Such an approach to  communication in markets using random graphs was
first suggested in the economics literature by Kirman      \cite{kirman1} 
to study the properties of the core
of a large economy.
Random graphs have also been used in the context of multilateral
matching in search problems by Ioannides      \cite{ioannides1}.
 A good review of the applications of random graph theory
in economic modeling is given in        \cite{ioannides2}.

The properties of large random graphs in the $N\to \infty$ limit were first studied
 by Erd\"os and Renyi      \cite{erdos}. An extensive review of mathematical results on random graphs is given in      \cite{bollobas}. The main resullts of the
combinatorial approach 
are given in Appendix 1.
One can show      \cite{bollobas} that for $c=1$ the probability density for the cluster size
 distribution decreases asymptotically as a power law:

$$
P(W) \mathop{\sim}_{W\to\infty} \frac{A}{W^{5/2}}
$$

while for values of $c$ close to and smaller than 1 ( $0 < 1-c << 1$), the cluster size distribution is cut off by an exponential tail:

\begin{eqnarray}
P(W)  \mathop{\sim}_{W\to\infty} \frac{A}{W^{5/2}} \exp(-\frac{(c-1)W}{W_0}) \label{tail}
\end{eqnarray}

For c=1, the distribution has an infinite variance while for $c < 1$ the variance
 becomes finite because of the exponential
tail.  In this case  the average size of a coalition is of order $1/(1-c)$
and the average number of clusters is then  of order $N(1- c/2)$.

Setting the coordination parameter $c$ close to 1 means that each agent tends to establish a link with one other agent, which can be regarded as a reasonable assumption. This does not rule out the formation of large coalitions through
successive binary links between agents but prevents a single agent from forming
multiple links, as would be the case in a centralized communication structure where one agent 
(the "auctioneer") is linked to all the others.
As argued by Kirman      \cite{kirman1}  the presence of a Walrasian auctioneer 
corresponds to such 'star-like', centralized, communication structures.
We are thus excluding such a situation by construction: we are interested
in a market where information is distributed and not centralized, which
corresponds more closely to the situations encountered in  real markets. 
More precisely, the local structure of the market may be characterized by the following result
     \cite{bollobas}: in the limit $N\to\infty$, the number $\nu_i$ of neighbors
of a given agent $i$ is a Poisson random variable with parameter $c$:
\begin{eqnarray}
P(\nu_i = \nu) &=& e^{-c}\frac{c^{\nu}}{{\nu}!} 
\end{eqnarray}

A Walrasian auctioneer $w$ would be connected to every other agent:
 $\nu_w= N-1$. The probability for having a Walrasian auctioneer is therefore given by $N P(\nu_w= N-1)$ which goes to zero when $N\to \infty$.

A given market cluster is characterized by its size $W_\alpha$ and
its 'nature' i.e. whether the members are buyers or sellers. This is
specified by a variable  $\phi_\alpha \in \{ -1,0,1\}$.
It is reasonable to assume that $W_\alpha$
and   $\phi_\alpha$  are independent
random variables: the size of a group does not influence its decision whether to buy or sell.
The variable  $X_\alpha= \phi_\alpha W_\alpha$ 
is then symmetrically distributed  with a mass of $1-2a$ at the origin.
Let  

\begin{eqnarray}
F(x) &=&  P(X_\alpha \leq x | X_\alpha \not= 0) \label{F}
\end{eqnarray}
 
Then the distribution of $X_\alpha$ is given by
\begin{eqnarray}
G(x) &=& P(X_\alpha\leq x) = (1-2a) H(x) + 2a F(x) \label{G}
\end{eqnarray}

where $H$ is a unit step function at 0 (Heaviside function).
We shall assume that $F$ has a continuous density, $f$.
$f$ then
decays
asymptotically  as in (\ref{tail}):
 
\begin{eqnarray}
 f(x)  \mathop{\sim}_{|x|\to\infty}
\frac{A}{|x|^{5/2}} e^{ \frac{-(c-1)|x|}{W_0}} 
\end{eqnarray}

The expression for the price variation 
$ { \Delta x}$ 
therefore reduces to a sum of $n_c$ {\it iid} random variables $X_\alpha, \alpha=1..n_c$ with heavy-tailed distributions as in (\ref{tail}):

$$  \Delta x = \frac{1}{\lambda} \sum_{\alpha=1}^{n_c} X_\alpha$$

Since the probability density of $X_\alpha$ has a finite mass $1-2a$ at zero, only a fraction
$2a$ of the terms in the sum (\ref{somme}) are non-zero; the number of non-zero terms
in the sum is of order $2a \overline{n_c} \sim 2a N (1- c/2) = N_{order} (1-c/2)$ where $N_{order} =
2a N$ is the average number
 of market participants who actively trade in the market
during a given period. For example, $N_{order}$ can be thought of as the number of
orders received  during the time period $[t,t+1]$ if we assume that different orders correspond 
to net demands, as defined above, of different clusters of agents. For a time period of, say, 15 minutes
on a liquid market such as NYSE, $N_{order} = 100 - 1000$ is a typical order of magnitude.

The distribution of the price variation $\delta x$ is then given by 
\begin{eqnarray}
P ( \Delta x =  x )  &=& \sum_{k=1}^{N}
P(n_c = k)
\sum_{j=0}^{k} \left( \begin{array}{c} k \\ j \end{array}\right) (2a)^j (1-2a)^{k-j} f^{\otimes j} (\lambda x) \label{convolution}
\end{eqnarray}

where $\otimes$ denotes a convolution product, $n_c$ being the number of clusters.
The above equation enables us to calculate the moment generating functions ${\cal{F}}$
of the aggregate excess demand $D$  in terms of $\tilde{f}$ (see Appendix 3 for details):

\begin{eqnarray}
{\cal{F}}(z)   \mathop{\sim}_{N\to\infty} \exp [N_{order}(1-\frac{c}{2}) (\tilde{f}(z) -1) ]  \label{laplace}
\end{eqnarray}

The moments of
$D$ (and those of $\Delta x$) may now be obtained through a Taylor expansion
of  Eq.(\ref{laplace}) (see Appendix 4 for details). The calculation of the variance and the fourth moment yields:
\begin{eqnarray}
\mu_2(D) &=& N_{order}(1-\frac{c}{2}) \mu_2(X_\alpha)\\
\mu_4(D) &=& N_{order}(1-\frac{c}{2}) \mu_4(X_\alpha) + 3 N_{order}^2(1-\frac{c}{2})^2 \mu_2(X_\alpha)^2
\end{eqnarray}
An interesting quantity  is the kurtosis of the  asset returns which, in our model, is equal
to the kurtosis of  excess demand $\kappa(D)$: 
$$
\kappa(D) = \frac{\mu_4(X_\alpha)}{N_{order}(1-\frac{c}{2}) \mu_2(X_\alpha)}
$$
The moments $\mu_j(X_\alpha)$ may be obtained may be obtained by an expansion in 1/N where N
is the number of agents in the market (see Appendix 2). 
Substituting their expression on the above formula yields the kurtosis $\kappa(D)$ as
a function of $c$ and the order flow:
\begin{eqnarray}
\kappa(D) =  \frac{2c+1}{N_{order}(1-\frac{c}{2}) A(c) (1-c)^3} \label{kurtosis}
\end{eqnarray}

where $A(c)$ is a normalization constant with a value close to 1 defined in Appendix 2, tending to a finite limit
as $c\to1$. 
This relation may be interpreted as follows: a reduction in the volume of the order flow results
in larger price fluctuations, characterized by a larger excess kurtosis. 
This result corresponds to the well known fact that large price fluctuations are more likely to occur
in  less active markets, characterized by a smaller order flow. It is also consistent
with results from various  market microstructure models where a  larger order flow enables easier
regulation of supply and demand by the market maker. It is interesting that we find the same
qualitative feature here although we have not explicitly integrated a market maker 
in our model.
This result should be compared to the observation in      \cite{engle91} that, even after accounting for
heteroskedasticity, the conditional distribution of stock returns for small firms is higher
than that of large firms. Small firm stocks being characterized by a smaller order flow $N_{order}$,
this observation is compatible with our results.

More importantly, Eq. (\ref{kurtosis}) shows that the kurtosis can be very large {\it even if
the number of orders is itself large}, provided $c$ is close to 1. 
Since $A(1)$ is close to 1/2, one finds that even for $c=0.9$ and $N_{order}=1000$, the kurtosis 
$\kappa$ is still of order 10, as observed on very active markets 
on time intervals of tens of minutes. Actually, one can show that provided $2 a N$ 
is not too large, the asymptotic behaviour of $P(\Delta x)$ is still of form given by Eq. (\ref{tail}).
 This model thus leads naturally to the value of $\mu=3/2$, close to the
value observed on real markets.
Of course, the value of $c$ could itself be time dependent. 
For example, herding tendency tends to be stronger during periods of uncertainty, 
leading to an increase in the kurtosis. When $c$ reaches one, a finite fraction of the market shares
simultaneously the same opinion and this leads to a crash. An interesting extension of the model would be one in which the time evolution
of the market structure is explicitly modeled, and the possible feedback effect of the 
price moves on the behavior of market participants.

\section{Discussion}

We have exhibited a model of a stock market which, albeit its simplicity, 
gives rise to a non-trivial probability distribution for aggregate excess demand and
stock price variations,
similar to empirical distributions of asset returns. 
Our model
illustrates the fact that while a naive market model in which agents do not interact
with each other would tend to give rise to normally distributed aggregate fluctuations, taking into account interaction between market participants
through a rudimentary 'herding' mechanism gives a result which is quantitatively
comparable to empirical findings on the distribution of stock market returns.
 
One of the interesting results of our model is that it predicts a relation between
the fatness of the tails of asset returns as measured by their excess kurtosis and
the degree of herding among market participants as measured by the parameter $c$.
This relation is given by Eq.(\ref{kurtosis}) .

Although we implicitly assumed that $t$ represents chronological time, one could formulate
the model by considering $t$ as ``market time", leading to a subordinated process in real
time as in      \cite{clark}, with the difference that the underlying process will not be a 
Gaussian random walk. 

Our model raises several interesting questions.
 As remarked above, the value of
$c$ is specified as being less than, and close to 1. ``Fine-tuning" a parameter to a certain value
may seem arbitrary unless one can justify such an  assumption. 
An interesting extension of the model would be one in which the time evolution
of the market structure is explicitly modeled in such a way that the parameter
$c$ remains in the critical region (close to 1).
 
One approach to this problem is via the concept 
 of "self-organized criticality", introduced by
Bak {\it et al}      \cite{bak87}: certain dynamical
systems generically evolve to a  state where the  parameters converge to the
critical
values leading to scaling laws and heavy-tailed distributions for the quantities modeled. This state is reached asymptotically and is an attractor for the
dynamics of the system.
Bak, Chen, Scheinkman and Woodford      \cite{bak93} present a simple model of an economic system presenting self-organized criticality.

Note however that, for the above results to hold, one does not need to adjust  $c$ to a
critical value : it is sufficient for $c$ to be within a certain range of values. As noted above, 
when $c$ approaches 1 the clusters become larger and larger and a giant
coalition appears when $c\geq1$. In our model the activation of such a cluster
 would correspond to a market crash (or boom).
In order to be realistic, the dynamics of $c$ should be such that
the crash (or boom) is {\it not} a stable state and the giant cluster disaggregates shortly
after it is formed: after a short period of panic, the market resumes 
normal activity. In mathematical terms, one should specify the dynamics of $c(t)$
such that the value
$c=1$ is  'repulsive'.  This can be achieved by introducing a feedback effect of prices
on the behavior of market participants: a nonlinear coupling between can lead to a control mechanism
maintaining  $c$ in the critical region.

Yet another interesting dynamical specification compatible with our model
is obtained by considering agents with "threshold response".
Threshold models have been previously considered as possible origins for collective
phenomena in economic systems      \cite{granovetter83}.
One can introduce heterogeneity by allowing the individual threshold
$\theta_i$ to be  random variables: for example one may assume the $\theta_i$s to be
{\it iid} with a standard deviation $\sigma(\theta)$. A simple way to introduce
interactions among agents is through  an aggregate variable:
each agent observes the aggregate excess demand $D(t)$ given by Eq.(\ref{demand}) or eventually $D(t) + E(t)$, where $E$ is an exogeneous variable.
Agents then evolve as follows:
at each time step, an agents changes  its market position $\phi(t)$ 
(``flips" from long to short or vice versa)
if
the observed signal $D(t)$ crosses his/her threshold $\theta_i$. 
Aggregate fluctuations  can then occur through cascades or ``avalanches"
corresponding to the flipping of market positions of groups of agents.
This model has been studied in the context of physical systems by
Sethna {\it et al}      \cite{sethna97} who have shown that for a fairly 
wide range
of values of $\sigma(\theta)$ one observes aggregate fluctuations
whose distribution has power-law behavior with exponential tails, as in Eq.(\ref{tail}).

These issues will be adressed in a forthcoming work. 
 
\vskip 0.5cm

\bibliographystyle{unsrt}

$\bullet$ {\bf Appendices:}\\

Unless specified otherwise, $f(N,c) \sim g(N,c)$
means 
$$
\frac{f(N,c)}{g(N,c)} = 1 + \mathop{o}_{N\to\infty}(1)
$$
uniformly in $c$ on all compact subsets of ]0,1[.

$\bullet$ {\bf Appendix 1: some results from random graph theory}\\

In this appendix we will review some results on
asymptotic properties of large random graphs. 
Proofs for most of the results  may be found in
     \cite{erdos} or      \cite{bollobas}.

Consider $N$ labeled points $V_1,V_2,...V_N$, called {\it vertices}.
A link (or edge) is defined as an unordered pair \{i,j \}.
A graph is defined by a set $V$ of vertices 
and a set $E$ of edges. Any two vertices may either be linked by one edge or not be linked 
at all. In the language of graph theory, we will consider
non-oriented  graphs without parallel edges.
We shall always denote the number of vertices by $N$.
A {\it path} is defined as a finite sequence of links such that every two
consecutive edges and only these have a common vertex. Vertices along a path
may be labeled in two ways, thus enabling to define the extremities of the path.
A graph is said to be connected if any two vertices $V_i, V_j$ 
are linked by  a path i.e. there exists
a path with  $V_i$ and $V_j$ as extremities.
A {\it cycle} (or loop) is defined  as a path such that the extremities coincide.
A graph is called a {\it tree} if it is connected and if none of its subgraphs is a cycle. 
A graph is called acyclic if all its subgraphs are trees.

Consider now a graph built by choosing, for each pair of vertices $V_i, V_j$ 
whether to link them or not through a random process, the probability for selecting any given
edge being $p>0$, the decisions for different edges being independent. A graph
obtained by such a procedure is termed a  {\it random graph} of type ${\cal G}(N,p)$ in the
notations of      \cite{bollobas}\footnote{This definition corresponds to random graphs of type
$\Gamma_{n,N}^{**}$ in      \cite{erdos} (see      \cite{erdos},p. 20).}.

In the following, we will be specifically interested in the case $p=c/N$.
Various graph-theoretical parameters of such graphs are random variables
whose distributions only depends on $N$ and $c$. We shall be particularly interested
in the properties of large random graphs of this type i.e. ${\cal G}(N,c/N)$ in the
limit $N\to\infty$.
 
The following results have been shown by Erd\"os and Renyi     \cite{erdos} and Bollobas      \cite{bollobas}:
If $c<1$ then in the limit $N\to\infty$ all point of the random graphs belong to trees except
for a finite number $U$ of vertices which belong to unicyclic components. 
Moreover, the probability of a vertex belonging to a cyclic component tends to zero as
$N^{-1/3}$. For describing the structure of large random graphs for $c<1$ it is therefore
sufficient to account for vertices belonging to trees; cyclic components do not
essentially modify the results, except when $c=1$.

More precisely (     \cite{bollobas}, Theorem  V.22)
\begin{eqnarray*}
\overline{U} &\sim &\frac{1}{2}\sum_{k=3}^{\infty} (ce^{-c})^k \sum_{j=0}^{k-3} \frac{k^j}{j!}\\
\sigma^2(U) &\sim &\frac{1}{2}\sum_{k=3}^{\infty} k(ce^{-c})^k \sum_{j=0}^{k-3} \frac{k^j}{j!}\\
\end{eqnarray*}

The above expressions are valid for $c\neq 1$.

$\bullet${\bf Appendix 2: Distribution of cluster sizes in a large random graph}

Let $p_1(s)$ be the probability for a given vertex to belong to
a cluster of size $s$ in the $N\to\infty$ limit.
The moment generating function $\Phi_1$
of the $p_1$ is defined by :
$$
\Phi_1(z) = \sum_{s=1}^{\infty} e^{sz} p_1(s)
$$

We shall now proceed to derive a functional equation verified by $\Phi_1$
in the large $N$ limit 
when the effect of loops (cycles) are neglected.

Let $p_N^1(s)$ be the corresponding probability in a random graph with $N$
vertices. Adding a new vertex to the graph will modify the pattern of links,
the probability of $k$ new links from the new vertex to the old ones being
$(c/N)^k (1-c/N)^{N-k} \left( \begin{array}{c} N \\ k \end{array}\right)$.
As shown above (Appendix 1) the probability of creating a cycle tends to zero
for large $N$.  The constraint that no new 
cycles are created by the new links imposes that the $k$ links
are made to vertices in $k$ different clusters of sizes. If $s_1,s_2,...s_k$ are the sizes of
these clusters, the new links will create a  new cluster of size $s_1 + s_2 +... +s_k + 1$. 

$$
p_{N+1}^1 (s) = \sum_{k=1}^{N} \sum_{s_1,...,s_k = 1}^{N}
\left( \begin{array}{c} N \\ k \end{array}\right)
(\frac{c}{N})^k (1-\frac{c}{N})^{N-k} \delta (s_1 + s_2 +... +s_k + 1 - s )
 p_N^1 (s_1) p_N^1 (s_2)... p_N^1 (s_k)
$$
Multiplying both sides by $e^{sz}$ and summing over $s$ gives:
$$
\Phi_1(z,N+1) = e^z [ 1 + \frac{c}{N} + \Phi_1(z,N)\frac{c}{N} ]^N 
$$
which gives 
 in the large $N$ limit:

$$
\Phi_1(z) = e^{z+ c(\Phi_1(z) -1)}
$$

from which various moments and cumulants may be calculated recursively.
The  distribution of clusters sizes  $p(s)$  is then given by
$$
p(s) = A(c) \frac{p_1(s)}{s}
$$
where $A(c)$ is a normalizing constant defined such that $\int p(s) ds = 1$.
\vskip 0.5cm

$\bullet$ {\bf Appendix 3: Number of clusters in a large random graph}\\

Let $n_c(N)$ be the number of clusters ( connected components) in a random 
graph of size N defined as above. $n_c$ is a random variable whose characteristics depend on $N$ and the parameter $c$.
In this section we will show that $n_c$ has an asymptotic normal distribution 
when $N\to\infty$ and that for large $N$, the j-th cumulant $C_j$ of 
of $n_c$ is given by:

$$
C_j \mathop{\sim}_{N\to\infty} \frac{(-1)^j N c}{2}
$$

From a  well known generalization of Euler's theorem in graph theory 

$$
l(N) - N + n_{c}(N) = \chi(N)
$$

where $\chi(N)$ is the number of independent cycles and $l(N)$ the number of links. 
This implies in turn 
$$
\overline{n_c(N)} = N (1- \frac{c}{2}) + O(1)
$$
We shall retrieve this result below, and proceed to calculate higher moments via 
an approximation.
Define the moment generating function for the variable $n_c(N)$ to be

$$
\Phi_N(z,c) = \overline{e^{n_c z}} = \sum_{k=1}^{N} P_{N,c}(n_c = k) e^{kz}
$$

The j-th moment of $n_c$ is then given by:
$$
\overline{{n_c}^j} = \frac{\partial^j\phi_N}{\partial z^j}(0,c)
$$
Let us also consider the cumulant generating function $\Psi$ defined by
$\Phi (z) = \exp \Psi (z)$.
The {\it j-th cumulant} of the distribution of $n_c$ may then be calculated as

$$
C_j (N,c)= \frac{\partial^j\Psi_N}{\partial z^j}(0,c)
$$

We will now establish an approximate recursion relation between
$\Phi_N$ and $\Phi_{N+1}$. Take a random graph of size $N$, the probability
of a link between any two vertices being $p= c/N$. In order to obtain
a graph with $N+1$ vertices, add a new vertex and choose randomly the links 
between the new vertex and the others. Note
that since in a graph of size $N$ the link probability is $p_N = c/N$, our new graph
will correspond to a graph of size $N+1$ with parameter $c'=c(N+1)/N$
so that the link probability is $c'/(N+1) = c/N$.
We will assume that the probability
of two links being made to the same cluster is  negligible i.e. that
no cycles are created by the new links, which is a reasonable approximation given the
results in Appendix 1. In this case, each $k$ links emanating from the new vertex will diminish the number of clusters by $k-1$, giving the following
recursion relation:

$$
P_{N+1,c'}(n) = (1-\frac{c}{N})^N P_{N+1,c'}(n +1) + \sum_{k=1}^{N}
\left( \begin{array}{c} N \\ k \end{array}\right)
(\frac{c}{N})^k (1-\frac{c}{N})^{N-k} P_{N+1,c'}(n+k-1)
$$
Multiplying each side by $ e^{nz}$ and summing over $n=1..N$ gives:

$$
\Phi_{N+1}(z,c') = e^z \Phi_N(z,c) [ 1 + \frac{c}{N}(e^{-z}-1)]^{N}
$$
 or, in terms of the cumulant-generating function $\Psi_N$:

$$
\Psi_{N+1}(z,c(1+\frac{1}{N})) = z + \Psi_N(z,c) + \ln ( 1 + \frac{c}{N}(e^{-z}-1)]^{N} \eqno(*)
$$

When $N\to\infty$, an first order expansion in $\frac{1}{N}$ gives:

$$
\Phi_{N+1}(z,c) + \frac{c}{N} \partial _2\Phi_{N+1}(z,c) =
e^z \Phi_N(z,c) \exp ({c(e^{-z}-1}) [1- \frac{c^2(e^{-z}-1)^2}{N}] + o(\frac{1}{N}) \eqno(**)
$$
where $\partial _2$ denotes a partial derivative with respect to the second variable. The second term on the left hand side stems from the
expansion  in the variable $c'=c(1+ 1/N)$  and reflects the fact that
the probability for a link has to be renormalized when going from a N-graph
to a N+1 - graph. 

By taking successive partial derivatives of (*) and (**) with respect to $z$ one can then
derive recursion relation for the moments and cumulants of $n_c$.
Let us first retrieve the result given in appendix one for $\overline{n_c}$.
Define $\gamma(c)$ such that
 $$\overline{n_c}= \frac{\partial\phi_N}{\partial z}(0,c) = \gamma_1(c)N + O(1)$$
Substituting in (*) yields a simple differential equation for $\gamma_1$:
$$
\gamma_1(c) + c {\gamma_1} '(c) = 1 -c 
$$
 whose solution is 
$\gamma_1(c) = (1- c/2) $ i.e. 
$$
\overline{n_c}= (1-\frac{c}{2}) N + O(1) \qquad \frac{\overline{n_c}}{N} \to 1-\frac{c}{2}
$$
Let us now derive a similar relation for the variance $\sigma^2(N,c)= var({ n_c})$. Let
$$
\sigma^2(N,c) = {\gamma_2}(c) N + O(1)
$$
By taking derivatives twice with respect to $z$ in (*) and setting $z=0$
one obtains, up to first order in $1/N$:
$$
\gamma_2(c) +  c {\gamma_2} '(c) = c
$$
whence 
$\gamma_2(c) = c/2 $
By calculating the j-th derivative in (*) with respect to $z$ one can 
derive in the same way an asymptotic expression for the $j$-th cumulant of $n_c$:
$$
C_j \mathop{\sim}_{N\to\infty} \frac{(-1)^j N c}{2}
$$

Note that the asymptotic forms for cumulants of $n_c$ are identical to those of
 a random variable $Z$ with the following distribution:

\begin{eqnarray*}
P(Z =k) = \frac{(\frac{Nc}{2})^{N-k}}{(N-k)!} e^{-\frac{Nc}{2}}
\end{eqnarray*}

i.e. $N - Z$ is a 
Poisson variable with parameter $Nc/2$. Without rescaling, this distribution becomes
degenerate in the large N limit. Nevertheless for finite $N$ both $\Phi_N$ and $\Psi_N$
are analytic functions of $z$ in a neighborhood of zero.
Consider now the rescaled variable:
$$
Y_N = \frac{n_c - N(1- \frac{c}{2})}{\sqrt{Nc/2}}
$$
$Y_N$ has zero mean and unit variance and its higher cumulants tend to zero:

$$  \forall j \geq 3, C_j( Y_N ) \mathop{\to}_{N\to\infty} 0 
$$ 
The standard normal distribution is the only distribution with zero mean, unit variance and
zero higher cumulants.
Under these
conditions, one can show      \cite{feller} that the convergence of the cumulants
implies  convergence in distribution:
 
$$ \frac{n_c - N(1- \frac{c}{2})}{\sqrt{Nc/2}} \mathop{\sim}_{N\to\infty} {\cal N}(0,1)$$

$\bullet$ {\bf Appendix 4:  distribution of aggregate excess demand }\\

In this appendix we derive an equation for the generating function ofthe variable $\Delta x$  which represents in our model
the one-period return of the asset. The relation between $\Delta x$ and other
variables of the model is given  by equation (5):
$$
{\Delta x} = \frac{1}{\lambda} \sum _{\alpha=1}^{n_c} W_{\alpha} \phi_{\alpha} = 
\frac{1}{\lambda} \sum _{\alpha=1}^{n_c} X_{\alpha}
$$
where $n_c$ is the number of clusters or trading group i.e. the number of
connected components of the random graph in the context of our model. $n_c$ is
itself a random variable, whose cumulants are known in the $N\to\infty$ limit (see Appendix 3).
As for the random variables $X_\alpha$, their distribution is given by Eq.(8)
 
\begin{eqnarray*}
P ( \Delta x =  x )  & = & \sum_{k=1}^{N}
P(n_c = k)
\sum_{j=0}^{k} \left( \begin{array}{c} k \\ j \end{array}\right) (2a)^j (1-2a)^{k-j} f^{\otimes j} (\lambda x) \\
\end{eqnarray*}

In order to calculate this sum, let us  introduce the moment generating functions
for $\Delta x$ and $X_{\alpha}'$:
$$
\tilde{f}(z) = \sum_{s}^{\ } f(s) e^{sz}\qquad {\cal{F}}(z)= \sum_{s}^{\ } P(\lambda\Delta x =  s) e^{sz}
$$

Multiplying the right hand side of the equation above  by $e^{\lambda x z}$ and summing over $s= \lambda x$ yields:

 \begin{eqnarray*}
{\cal{F}}(z) &=&
\sum_{k=1}^{N} P(n_c = k) [ 1-2a + 2a \tilde{f}(z)]^k\\
&=& \Phi [\ln (1 +  2a (\tilde{f}(z) - 1) )] \\
&=& \exp [\Psi (\ln (1 +  2a (\tilde{f}(z) - 1) )  ]
\end{eqnarray*}

where $\Psi(z)$ is the cumulant generating function of the number of
clusters defined in appendix 3.  $\Psi$ is an analytic function whose series expansion
is given by the cumulants of $n_c$
$$
\Psi(z) = Nz + \frac{Nc}{2} \sum_{k=1}^{\infty} \frac{(-z)^j}{j!} = Nz + \frac{Nc}{2}(e^{-z}-1)
$$

 one can evaluate the above sum in the large N limit as
 \begin{eqnarray*}
{\cal{F}}(z) &=&\exp [\Psi (\ln (1 +  2a (\tilde{f}(z) - 1) )  ]\\
&=&\gamma^N \exp [\frac{Nc}{2}(\frac{1}{\gamma}-1)]
\end{eqnarray*}
where 
$$
\gamma = [1 + 2a ({\tilde{f}}(z) - 1) ] 
$$

Recall that $2a$ corresponds to the fraction  of agents who are active in the market
 in a given period. Therefore
$ 2a N$ is
the average number  of buy and sell orders sent to the market in one period. We shall choose
$a(N)$ such that in the limit $N\to\infty$ the number of orders 
has a finite limit, which we will denote by $N_{orders}$ : 
  $ 2a N \to N_{orders} $. More precisely if
we assume that 
$2a = N_{orders}/N + o(1/N)$ then
\begin{eqnarray*}
\gamma^N &=& \exp[N_{orders}({\tilde{f}}(z) - 1)] + O(\frac{1}{N})\\
 (\frac{1}{\gamma}-1) &=& -\frac{N_{orders}({\tilde{f}}(z) - 1)}{N} + o(\frac{1}{N}) 
\end{eqnarray*}
in the above expression  gives:
\begin{eqnarray*}
{\cal{F}}(z)&=&\gamma^N \exp [\frac{Nc}{2}(\frac{1}{\gamma}-1)] \\
&=&\exp[N_{orders}({\tilde{f}}(z) - 1)] \exp[\frac{-C N_{orders}}{2} ({\tilde{f}}(z) - 1)] \\
&=&\exp [ N_{orders}(1-\frac{c}{2}) ({\tilde{f}}(z) - 1) ] + O(\frac{1}{N})  
\end{eqnarray*}

One finally obtains:

$$
{\cal{F}}(z)  \simeq  \exp [N_{order}(1-\frac{c}{2}) (\tilde{f}(z) -1) ]
$$

Let us now examine the implication of the above relation for  the moments of
$D$ and $\Delta x$. Expanding both sides in a Taylor series yields:

\begin{eqnarray*}
\mu_2(D) &=& N_{order}(1-\frac{c}{2}) \mu_2(X_\alpha)\\
\mu_4(D) &=& N_{order}(1-\frac{c}{2}) \mu_4(X_\alpha) + 3 N_{order}^2(1-\frac{c}{2})^2 \mu_2(X_\alpha)^2
\end{eqnarray*}

which implies that the kurtosis $\kappa(D)$ of the aggregate excess demand is given by
$$
\kappa(D) = \frac{\mu_4(X_\alpha)}{N_{order}(1-\frac{c}{2}) \mu_2(X_\alpha)}
$$

\end{document}